\documentclass[10pt]{article}

\usepackage{amsmath}
\usepackage{graphicx}
\usepackage{compozition} \setdefaultfont{\mathnormal} 

\title{\textbf{Proposal to use Humans to switch settings in a Bell experiment}}
\author{Lucien Hardy\\
\textit{Perimeter Institute,}\\
\textit{31 Caroline Street North,}\\
\textit{Waterloo, Ontario N2L 2Y5, Canada}}
\date{}

\begin{document}

\maketitle

\begin{abstract}
In this paper I discuss how we might go about about performing a Bell experiment in which humans are used to decide the settings at each end.  To get a sufficiently high rate of switching at both ends, I suggest an experiment over a distance of about 100km with 100 people at each end wearing EEG headsets, with the signals from these headsets being used to switch the settings.

The radical possibility we wish to investigate is that, when humans are used to decide the settings (rather than various types of random number generators), we might then expect to see a violation of Quantum Theory in agreement with the relevant Bell inequality.  Such a result, while very unlikely, would be tremendously significant for our understanding of the world (and I will discuss some interpretations).

Possible radical implications aside, performing an experiment like this would push the development of new technologies.  The biggest problem would be to get sufficiently high rates wherein there has been a human induced switch at each end before a signal as to the new value of the setting could be communicated to the other end and, at the same time, a photon pair is detected.   It looks like an experiment like this, while challenging, is just about feasible with current technologies.
\end{abstract}

\section{Introduction}

In this proposal I discuss performing an experiment to test Bell's inequalities \cite{bell1964einstein} wherein humans are used to change the settings at the two ends.   The basic idea is that we perform a Bell experiment over a scale of about 100km and have, at each end, about 100 humans who intervene on the settings via electrical brain activity obtained by electrodes placed on their scalps to intervene on the settings (as is done in recording an electroencephalogram (EEG)).  We want to have a large number of cases where the setting has been changed by human interventions at both ends while a signal as to the new value of the setting cannot have yet reached the other side. We suggest using EEG brain activity (rather than, for example, pressing a button by hand) to minimize delays.  We need the experiment to be over a large distance scale and to have many humans at each end to get a sufficiently high rate that we could expect a significant effect.  Having more humans at each end would increase the rate. Making the experiment longer is also good as long as we have a high enough coincidence count rate that we can still get a significant effect.  The parameters suggested above (100km with 100 people at each end) may be insufficient or more than  sufficient for our purposes.

The radical possibility we wish to investigate is that when humans are used to decide the settings (rather than various types of random number generators) we might then expect to see a violation of Quantum Theory in agreement with the relevant Bell inequality.  Such a result, while very unlikely, would be tremendously significant for our understanding of the world. A violation of Quantum Theory under these circumstances would, of course, be very important in and of itself - it would teach us that the world was, after all, fundamentally local as well as having implications for determinism as we will discuss.  But the real importance of such a result would be the demonstration that humans have a special role when compared with computers, machines, random number generators \dots  As we will discuss later, a natural explanation of such a result would be that it demonstrates some sort of Cartesian mind-matter duality (though one could seek out other types of explanation).

Possible radical implications aside, performing an experiment like this would push the development of new technologies.  The biggest problem would be to get sufficiently high rates wherein there has been a human induced switch at each end before a signal as to the new value of the setting could be communicated to the other end and, at the same time, a photon pair is detected.  This would require us to distribute entangled pairs of systems at a high rate over a scale of, at least, kilometers and possibly hundreds of kilometers.  Additionally, we would have to develop fast electronics applicable to these kinds of experiments.  The objective of performing this experiment would act as a stretch goal pushing us beyond what we might otherwise attempt and building on what has already been experimentally achieved \cite{freedman1972experimental, aspect1982experimental, tittel1998violation, weihs1998violation, scheidl2010violation} without attempting to have human input into the switching. Even in the most likely scenario that Quantum Theory emerges unscathed, there could still be technological payoffs to performing such an experiment. Such possible technological payoffs include: (i) developing even more robust and higher rate entanglement distribution schemes over greater distances; and (ii) opening up the field of coupling human choices to quantum systems using EEG technology (the use of EEG signals to input human choices in computers and robotics is already an active area of research \cite{wolpaw2002brain}).   We can investigate possible applications of such technology.  In particular, there may be security advantages to coupling humans directly to the apparatus in quantum cryptography \cite{wiesner1983conjugate, bennett1984quantum, ekert1991quantum}.  In particular, in device independent cryptography \cite{barrett2005no, acin2006bell} in which we assume there cannot be signalling faster than the speed of light, there could be security payoff to implementing locality conditions with respect to human interventions.

\section{Previous discussion}

In 1989 I wrote two papers \cite{hardy1989proposal, hardy1989local} on the idea of using humans to choose the settings in Bell experiments.  These papers did not, of course, get past the referees (and this was before quant-ph on the arXiv). Such ideas were certainly too speculative at that time.  Nevertheless, my PhD supervisor, Euan Squires summarized my idea in his beautiful book \lq\lq Conscious mind in the physical world" \cite{squires1990conscious} (a physicist's take on the issue of consciousness):
\begin{quote}
An [...] idea being studied by a research student here in Durham, L Hardy, is that there might exist genuine free agents which are outside the physically determined world. Such free agents could be responsible for \lq\lq mind-acts" affecting the settings in the EPR experiment.  Assuming these are constrained by the Bell inequality, they would give rise to violations of quantum theory. (Experiments along these lines would be precise tests of a well defined type of dualism. Unfortunately, the time scales involved suggest they would be very difficult to perform).
\end{quote}
The free agents in question being, of course, humans.  In 1990, when Squires wrote these words, it was clear that such an experiment was well beyond available technology.  However, there now exist much more efficient sources of entangled systems and Bell experiments have been performed over kilometers (and even hundreds of kilometers).  Further, fast switching techniques have been developed.  It seems that, by now, an experiment like this could be achieved by sufficiently determined experimentalists.  For the first iteration of such experiments it would sufficient to attempt to implement human switching without also closing the detection efficiency loop hole.  More recently I wrote another paper on using humans to switch the settings \cite{hardy2017bell}. That paper focussed on deriving Bell inequalities with retarded settings (see Sec.\ \ref{sec:Bellretarded}) which would be useful in such an experimental test while this paper focuses on how to actually implement an experiment.

In 1990 John Bell published his Nouvelle Cuisine paper.  In this he considers a Bell-type experiment and says
\begin{quote}
Then we may imagine the experiment done on a such a scale, with the two sides of the experiment separated by a distance of order light minutes, that we can imagine these settings being freely chosen at the last second by two different experimental physicists, or some other random devices.
\end{quote}
Bell, however, did not appear to explore the idea that such an experiment would relate to the mind-matter duality debate.  He says
\begin{quote}
I would expect a serious theory to permit \lq\lq deterministic chaos" or \lq\lq pseudorandomness", for complicated subsystems (e.g.\ computers) which would provide variables sufficiently free for the purpose at hand.
\end{quote}
Experimental physicists, from Bell's point of view, are an example of such subsystems.

In the last decade the idea of using humans to do the switching in Bell experiments has been mentioned in passing a number of times.  Aside from \cite{hardy1989proposal, hardy1989local, hardy1989local, hardy2017bell} there has been no discussion about the importance of such an experiment for the issue of mind-matter duality - one exception being the recent paper, \lq\lq Quantum and Qualia" by Adrian Kent \cite{kent2016quanta} who mentions this kind of experiment in the context of a broader discussion of about consciousness.  As described in Sec.\ \ref{sec:abellexperiment} and Sec.\ \ref{sec:feasibility}, Weihs \emph{et al}  \cite{weihs1998violation} performed the first experiment in which random number generators were used to choose the settings.  At the end of the paper they say
\begin{quote}
Further improvements, e.g. having a human observers choose the analyzer directions would again necessitate major improvements of technology as was the case in order
to finally, after more than 15 years, go significantly beyond the beautiful 1982 experiment of Aspect et al.
\end{quote}
The possibility of using experimental physicists to choose the settings is also mentioned in the Canary Islands paper \cite{scheidl2010violation} by Scheidl \emph{et al}.
More recently, Hensen \emph{et al} \cite{hensen2015loophole}, who performed one of three recent experiments that simultaneously closed the detector efficiency and switching loopholes say, at the end of their paper
\begin{quote}
Even so, our loophole-free Bell test opens the possibility to progressively bound such less conventional theories: by increasing the distance between $A$ and
$B$ (testing e.g. theories with increased speed of physical influence), using different random input bit generators (testing theories with specific free-will agents, e.g. humans).
\end{quote}
Very recently there was a public outreach initiative - The Big Bell Test - run by a number of experimental groups around the world \cite{BigBellTest2016}.  In this members of the public were encouraged to provide input over the internet which was used to switch the settings in various Bell experiments.  No attempt was made to impose locality conditions (these human choices were clearly in the backward light cone of both ends of the Bell experiment). However, the idea resonates with the ideas being discussed here.

Since I first attempted to publish ideas along these lines in the late 1980's attitudes have clearly changed so that there may be interest in doing this kind of experiment.  Further, the technology has developed tremendously so that such an experiment may be feasible.  Of course, if we are going to do an experiment of this sort it is worth being very careful up front as to what the idea is we are really trying to test and that the conditions for a genuine test are realized in any actual experiment.  


\section{The need for interventions to determine settings while particles are in flight}\label{sec:inflight}

When we derive Bell inequalities we assume that the outcome at ends $A$ and $B$ are given by some result functions
\begin{equation}
A(a, \lambda) ~~~~~~~~~~ B(b, \lambda)
\end{equation}
respectively.  Here $a$ and $b$ are the settings and $\lambda$ are the hidden variables.   It has long been appreciated that the settings in Bell experiments need to be chosen while the particles are in flight to ensure that the choice cannot be communicated to the other end by non-superluminal signals.

Consider an experiment in which the settings at each end are static.  In this case it is possible that, in the underlying physics, the setting is broadcast from each end to the other end so the outcomes at each end can depend on both settings. Then we would have result functions $A(a,b, \lambda)$ and $B(b,a,\lambda)$.  This would block the derivation of Bell inequalities and, indeed, we can easily construct a local model that reproduces Quantum Theory.

\begin{figure}
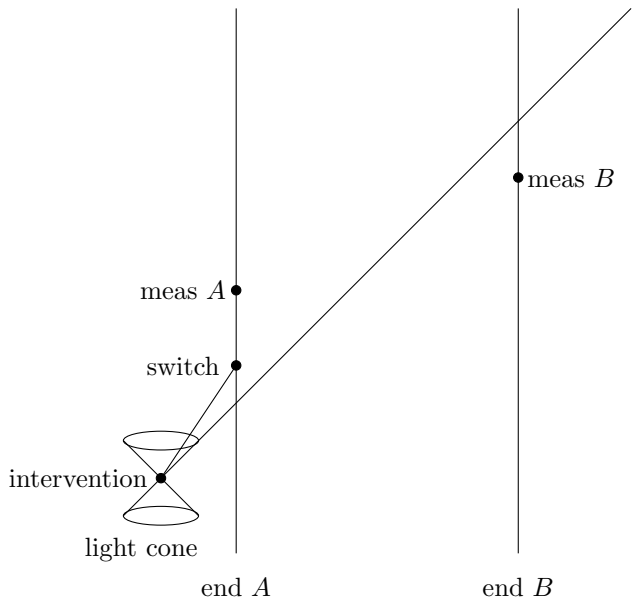

\begin{center}
\begin{Compose}{0}{0}
\thispoint{bL}{4,-4} \csymbol[0,-50]{\ensuremath{\text{end}~A}} \thispoint{tL}{4,25}   \jointbnoarrow{bL}{0}{tL}{0}
\thispoint{bR}{19,-4} \csymbol[0,-50]{\ensuremath{\text{end}~B}}\thispoint{tR}{19,25} \jointbnoarrow{bR}{0}{tR}{0}
\blackdot{int}{0,0} \csymbol[-125,0]{\ensuremath{\text{intervention}}}
\thispoint{LCr}{25,25} \thispoint{LCrm}{2,-2} \thispoint{LClm}{-2,-2} \thispoint{LCl}{-2,2}
\linebyhand[thin]{LCrm}{LCl} \linebyhand[thin]{LClm}{LCr}
\cellipse[thin]{LCplus}{2}{0.5}{0,2} \cellipse[thin]{LCminus}{2}{0.5}{0,-2}  \csymbol[-30,-50]{\ensuremath{\text{light cone}}}
\blackdot{switch}{4,6} \csymbol[-80,0]{\ensuremath{\text{switch}}}
\linebyhand[thin]{int}{switch}
\blackdot{measA}{4,10} \csymbol[-80,0]{\ensuremath{\text{meas}~A}}
\blackdot{measB}{19,16} \csymbol[80,0]{\ensuremath{\text{meas}~B}}
\end{Compose}
\end{center}
\caption{This figure shows an intervention changing the setting at end $A$ such that no signal carrying information about this intervention can reach end $B$.}
\label{fig:intervention}
\end{figure}

We could imagine having some machine decide the setting at each end. But in this case it is possible that the machines runs according to deterministic rules.  Then the earlier state of the machines can be communicated to the other end at non-superluminal speeds and from this earlier state the setting can be determined.  In this case we can still have $A(a, b,\lambda)$ and $B(b,a,\lambda)$ dependances.  What we need is an \emph{intervention} at each end that changes the setting from the value it would have taken had there been no intervention in a way that is spacelike separated from the measurement event at the other end (see Fig.\ \ref{fig:intervention}).

For clarity, it is worth mathematically elaborating this simple idea.  Let the state of machine at time $t$  be $\alpha_t$.  This state may consist of hidden variables that appear in the fundamental theory (that are not directly accessible to experimentalists). Further, this state can describe any physical systems that can locally influence the setting (so the term \lq\lq machine" is potentially broader than just referring to the box in the laboratory that appears to determine the setting).   According to our assumptions, in the absence of interventions, this state is given by some deterministic rules from the state at time $0$
\begin{equation}
\alpha_t = f_t(\alpha_0)
\end{equation}
We chose time $t=0$ to be the last time that a light speed signal can communicate the state of machine at end $A$ to the measurement event at end $B$.  If an intervention happens at a time $t'>0$ then
\begin{equation}
\alpha(t') \not= f_{t'}(\alpha_0)
\end{equation}
This is, simply, what we mean by an intervention. We can write the \emph{retarded setting} as
\begin{equation}
a_r = a(\alpha_0)
\end{equation}
This is the prediction as to what value the setting would take according to the last information available at end $B$.   We can have a result function $B(b, a_r, \lambda)$ for end $B$.  However, when there is an intervention we may have $a_r\not=a$ where $a$ is the actual setting.  Similar remarks apply for the state, $\beta_t$, of the machine at end $B$.   Hence, if we have interventions then we can recover Bell inequalities (actually, we can derive Bell inequalities with retarded settings as discussed in Sec.\ \ref{sec:Bellretarded}) which are violated by Quantum Theory.  The important point about these interventions is not so much that they are freely chosen but, rather, that they \lq\lq wrong foot" the attempt at the other end to predict the setting.  It could be that there is another deterministic function that can predict the interventions but if it is not the one used to calculate the retarded settings at the other end (in the supposed underlying physical model) then we can still expect a violation of the Bell inequalities in such a model.

The pertinent question now is what are suitable candidates for interventions of this type.  Here are some possibilities:
\begin{description}
\item[Random number generators:] By now there are many experiments using various types of random number generators including quantum random number generators \cite{weihs1998violation, scheidl2010violation, hensen2015loophole, giustina2015significant, shalm2015strong}
\item[Signals from distant galaxies:] It is possible, depending on what cosmological model one adopts, that distant galaxies have never been in causal contact. Hence, light from such galaxies would be a good candidate for interventions.
\item[Humans:] This is the possibility we discuss in this paper.
\end{description}
The problem with random number generators is that the underlying physics that describes them may actually be deterministic. Even quantum random number generators of the kind used in recent experiments may be governed by an underlying deterministic model.  Furthermore, very convincing experiments have now been performed in which Bell's inequalities were violated with random number generators.   Whether signals from distant galaxies are a good way to implement interventions really depends on what cosmological model one adopts.  However, it is generally believed that there are causally disconnected regions of the universe and so it is certainly it is worth pushing this direction of research.  One experiment has been performed using light signals from distant stellar sources in our galaxy \cite{handsteiner2017cosmic}.  This is taken to be the first step in a series of experiments that could use more and more remote cosmological systems to decide the settings. The last option, using humans, is the most radical.  We will discuss the interpretation of this in Sec.\ \ref{sec:mindmatter} - in particular we will consider that such interventions might be related to Cartesian mind-matter duality.

\section{Components of an experiment}

\begin{figure}[!b]
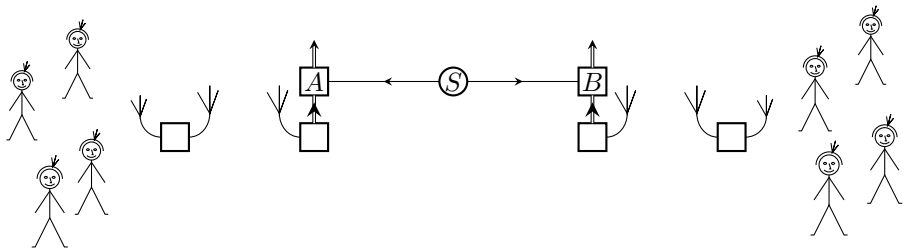

\begin{center}
\begin{Compose}[0.74]{-15}{0}
\Ccircle{S}{1}{0,0} \Csquare{A}{1}{-10,0}  \Csquare{B}{1}{10,0}
\csquare{Aset}{1}{-10, -4} \antena[0.8]{Aantset}{-12.5,-2.3} \joinlcnoarrow{Aset}{0}{Aantset}{-90}
\csquare{Bset}{1}{10,-4}   \antena[0.8]{Bantset}{12.5,-2.3}  \joinrcnoarrow{Bset}{0}{Bantset}{-90}
\csquare{Arec}{1}{-20, -4} \antena[0.8]{AantrecR}{-17.5, -2.3} \joinrcnoarrow{Arec}{0}{AantrecR}{-90} \antena[0.6]{AantrecL}{-22.5, -2.5} \joinlcnoarrow{Arec}{0}{AantrecL}{-90}
\csquare{Brec}{1}{20,-4}   \antena[0.8]{BantrecR}{17.5, -2.3}  \joinlcnoarrow{Brec}{0}{BantrecR}{-90} \antena[0.6]{BantrecL}{22.5, -2.5}  \joinrcnoarrow{Brec}{0}{BantrecL}{-90}
\thispoint{Aout}{-10,3} \thispoint{Bout}{10,3}
\joincr{S}{180}{A}{0} \joincl{S}{0}{B}{0}
\jointbdouble{Aset}{0}{A}{0} \jointbdouble{Bset}{0}{B}{0}
\jointbdoubleend{A}{0}{Aout}{0} \jointbdoubleend{B}{0}{Bout}{0}
\humanEEG[0.6]{-31,0}
\humanEEG[0.7]{-29,-7}
\humanEEG[0.65]{-26,-5}
\humanEEG[0.6]{-27,3}
\humanEEG[0.72]{27,-6}
\humanEEG[0.65]{26,1}
\humanEEG[0.68]{31,-4}
\humanEEG[0.6]{30, 4}
\end{Compose}
\end{center}
\caption{The proposed experiment.  This experiment has various components; (i) A long Bell experiment with electric input controlling settings at each end;
(ii) A large number of humans providing input at each end via EEG headsets; (iii) Radio frequency receivers and transmitters communicate the signal from the headsets to the switching devices at each end. In the figure we show a booster station at each end to receive the signal from the headsets and retransmit a composite signal that is used as input into the setting at each end.}
\label{fig:proposal}
\end{figure}

The basic proposal is that we perform a Bell experiment over some distance $d_\text{sep}$ with $N_A$ humans doing the switching for end $A$ and $N_B$  humans for end $B$ (see Fig.\ \ref{fig:proposal}).  These humans would wear EEG headsets and this electrical activity would be used to change the setting of the measurements at each end of the Bell experiment.  The humans do not need to be in the immediate vicinity of the given end.  The humans responsible for the switching at end $A$, for example, could be located some distance away (in the opposite direction from end $B$) with the switch signal transmitted via radio frequency to the switching device.

For a certain fraction, $\alpha$, of the coincidences collected, there will be a human induced switch at each end while no light speed signal can have transmitted this information to the other end.  We can imagine two strategies for analyzing the data: either (i) we can consider all the data and expect to see a shift in the violation of Bell's inequalities proportional to $-\alpha$ if there is a real effect; or (ii) we can consider a template that filters for certain types of of EEG activity that we suppose are associated with human interventions and test Bell's inequalities for those cases when the EEG signal satisfies this template at each end.

We will discuss each of the basic components of the proposed experiment.

\subsection{A Bell experiment}\label{sec:abellexperiment}

In a Bell experiment a source of entangled pairs of systems (system 1 and system 2) are sent to  two stations, $A$ and $B$, where measurements are made. At each station there is a choice between two (or more) settings for the measurements.  The source might, for example, be photons entangled in their polarizations.  In this case the measurement would be of polarization and the settings along one of two (or more) different angles.

For the purposes of the current proposal, we need to implement a Bell experiment that, under ordinary operation (without humans performing the switching), violates the Bell inequalities by a significant amount.  We require that this experiment is capable of fast switching between the two (or more) settings of the measurement at each end where these fast switches have an electrical input that determines the setting.

\begin{figure}
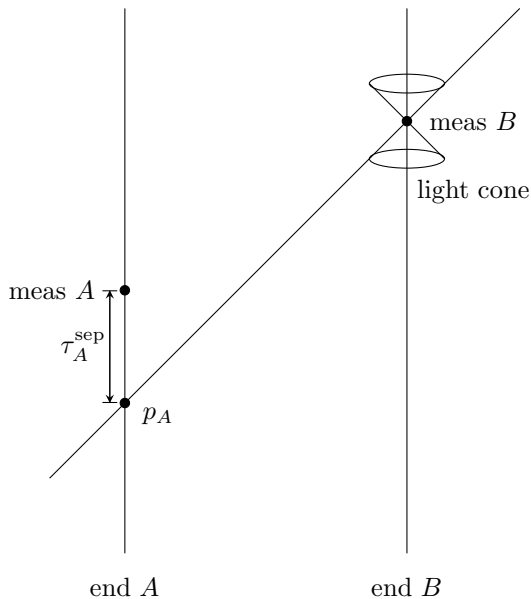

\begin{center}
\begin{Compose}{0}{0}
\thispoint{bL}{4,-4} \csymbol[0,-50]{\ensuremath{\text{end}~A}} \thispoint{tL}{4,25}   \jointbnoarrow{bL}{0}{tL}{0}
\thispoint{bR}{19,-4} \csymbol[0,-50]{\ensuremath{\text{end}~B}}\thispoint{tR}{19,25} \jointbnoarrow{bR}{0}{tR}{0}
\thispoint{int}{0,0}
\thispoint{LCr}{25,25} \thispoint{LCrm}{21,17} \thispoint{LClm}{17,17} \thispoint{LCl}{17,21}
\linebyhand[thin]{LCrm}{LCl} \linebyhand[thin]{0,0}{LCr}
\cellipse[thin]{LCplus}{2}{0.5}{19,21} \cellipse[thin]{LCminus}{2}{0.5}{19,17}  \csymbol[100,-50]{\ensuremath{\text{light cone}}}
\blackdot{measA}{4,10} \csymbol[-110,0]{\ensuremath{\text{meas}~A}}
\blackdot{measB}{19,19} \csymbol[100,0]{\ensuremath{\text{meas}~B}}
\linebyhand[thin, |-|]{3.2,4}{3.2,10} \linebyhand[thin, <->]{3.2,4}{3.2,10} \csymbol[-40,0]{\ensuremath{\tau_A^\text{sep}}}
\blackdot{pA}{4,4} \csymbol[50,-20]{\ensuremath{p_A}}
\end{Compose}
\end{center}
\caption{Definition of $\tau_A^\text{sep}$. We define $\tau_B^\text{sep}$ similarly.}
\label{fig:tauAsep}
\end{figure}

The two stations will be a certain distance, $d_\text{sep}=c\tau_\text{sep}$, apart.  It is is possible that, in the earth rest frame the measurement at end $A$ happens at a different time than the measurement at end $B$. This could happen if, for example, the source of the entangled systems is closer to one end (as in the Canary Islands experiment \cite{scheidl2010violation} discussed in Sec.\ \ref{sec:feasibility}).  Let $C_B$ be the backward light cone from the measurement event at end $B$ (see Fig.\ \ref{fig:tauAsep} for illustration of these remarks).  Let $p_A$ be the intersection point of $C_B$ with the world-line of station $A$.  We define $\tau_A^\text{sep}$ as the time interval, from the point $p_A$ to the measurement event at end $A$.  We define $\tau_B^\text{sep}$ similarly.  If the stations, $A$ and $B$, are in the same rest frame and the measurements happen simultaneously in that rest frame then $\tau_A^\text{sep}=\tau_B^\text{sep}=\tau_\text{sep}$.

Other key parameters of this experiment are: (i) the time, $T_\text{exp}$, required to obtain a violation of the Bell inequalities to some given significance; and (ii) the time delay from the time the switch signal is inputted into the fast switch till the time the measurement setting is changed at each end (these will contribute to the overall delay defined in Sec.\ \ref{sec:humaninput}).

There have been very many experiments to test Bell's theorem beginning with Freedman and Clauser's experiment \cite{freedman1972experimental} in 1972 which saw a violation of Bell's inequalities by 6 standard deviations.  The first experiment to switch the settings while the systems are in flight was performed by Aspect, Dalibard, and Roger in 1982 \cite{aspect1982experimental}.  A problem with this experiment was that the switching was periodic and, furthermore, the period was unfortunately chosen just such that the setting was back to its original value in the time taken for a signal to go from one end to the other \cite{zeilinger1986testing, hardy2017bell}.  The first experiment in which a random number generator was used to determine the settings was performed in in 1998 over 400m across the campus of the University of Innsbruck \cite{weihs1998violation}.  There have been many such experiments since then \cite{scheidl2010violation, hensen2015loophole, giustina2015significant, shalm2015strong}.  In particular, in 2010 an experiment was performed over a distance of 144km in the Canary Islands between La Palma and Tenerife using a free space link \cite{scheidl2010violation}.  There have also been some long experiments that did not attempt to choose the settings randomly. Of particular note is Gisin's team's experiment using optical fibers with the source in Geneva and the two ends in the neighbouring towns of Bellevue and Bernex (a distance of 10.9km apart) \cite{tittel1998violation}.

Very recently experiments have been performed that performed fast switching at the same time as also closing the fair sampling loophole \cite{hensen2015loophole, giustina2015significant, shalm2015strong}.  For our present purposes we do not propose attempting to close the fair sampling loophole (as important as this is to close, it does seem that nature would have to be especially conspiratorial to take advantage of it).  Future versions of this experiment might also attempt to close the fair sampling loophole and use humans to do the switching - this might be regarded as the ultimate Bell experiment.

So far, no experiment to date has used humans to determine the settings while the systems are in flight. Nevertheless, the performed experiments have many of the features that we will require and so they can be used to investigate the feasibility of the proposed experiment (as discussed in Sec.\ \ref{sec:feasibility}).

\subsection{Human input}\label{sec:humaninput}

\begin{figure}
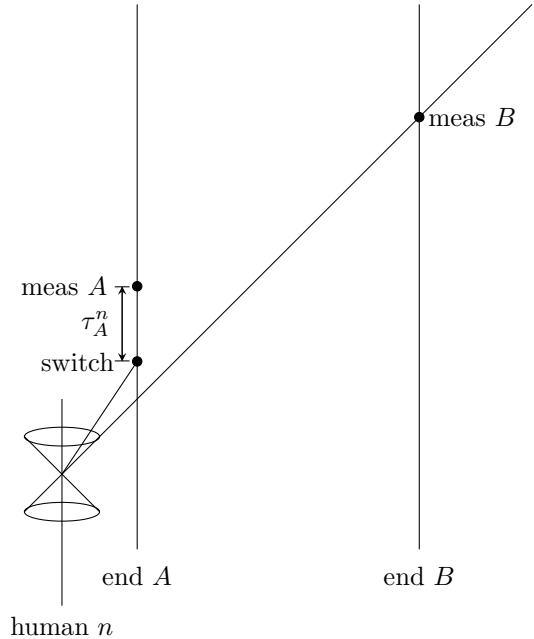

\begin{center}
\begin{Compose}{0}{0}
\thispoint{bL}{4,-4} \csymbol[0,-40]{\ensuremath{\text{end}~A}} \thispoint{tL}{4,25}   \jointbnoarrow{bL}{0}{tL}{0}
\thispoint{bR}{19,-4} \csymbol[0,-40]{\ensuremath{\text{end}~B}}\thispoint{tR}{19,25} \jointbnoarrow{bR}{0}{tR}{0}
\thispoint{bhumann}{0,-7} \csymbol[0,-30]{\ensuremath{\text{human}~n}} \thispoint{thumann}{0,4}   \jointbnoarrow{bhumann}{0}{thumann}{0}
\thispoint{int}{0,0}
\thispoint{LCr}{25,25} \thispoint{LCrm}{2,-2} \thispoint{LClm}{-2,-2} \thispoint{LCl}{-2,2}
\linebyhand[thin]{LCrm}{LCl} \linebyhand[thin]{LClm}{LCr}
\cellipse[thin]{LCplus}{2}{0.5}{0,2} \cellipse[thin]{LCminus}{2}{0.5}{0,-2} 
\blackdot{switch}{4,6} \csymbol[-90,0]{\ensuremath{\text{switch}}}
\linebyhand[thin]{int}{switch}
\blackdot{measA}{4,10} \csymbol[-110,0]{\ensuremath{\text{meas}~A}}
\blackdot{measB}{19,19} \csymbol[80,0]{\ensuremath{\text{meas}~B}}
\linebyhand[thin, |-|]{3.2,6}{3.2,10} \linebyhand[thin, <->]{3.2,6}{3.2,10} \csymbol[-40,0]{\ensuremath{\tau_A^n}}
\end{Compose}
\end{center}
\caption{Definition of $\tau_A^n$. The signal from the intervention to the switch may be at subluminal speeds.}
\label{fig:tauAn}
\end{figure}

We consider $N_A$ humans at end $A$ and $N_B$ humans at end $B$.  Each human would wear a EEG headset (to collect electrical brain activity).  The signals from these headsets would be transmitted (possibly via radio frequency) to an electronic device at each end that inputs a signal into the Bell experiment to switch the setting.  We could consider, instead, instructing the humans to press a button but this would operate at mechanical speeds and we would, correspondingly, require a Bell experiment on a much larger scale to hope to get useful results. Indeed, EEG analysis can predict human choices about one tenth of a second before buttons are pressed \cite{blankertz2002classifying}.  In this time a light speed signal will transverse several times the radius of the earth.  Since it is not practical to have two groups of humans separated by several times the radius of the earth we propose, instead, using EEG headsets.  A concern here is that we want be sure that EEG signals that are associated with human choices are not too delayed by the time they register in the EEG headset (i.e.\ we need to study how long it takes such signals to get from the part of the brain in which they originate to the outer surface of the skull).  This, however, appears not to be a problem.  According to \cite{fisch1999fisch} \lq\lq The transmission of a signal through a volume conductor occurs nearly at the speed of light. Therefore, similar appearing waveforms that occur in different or non-adjacent scalp locations without any difference in their time of appearance most likely arise from the same cortical generator."

We do not require that the humans involved deliberately and intentionally switch the settings to this or that value but rather that they engage in appropriate behaviour that we get a high rate of human interventions as defined in Sec.\ \ref{sec:neurologicalanalysis} (wherein we see certain features of the EEG signals that are conjectured to correspond to interventions - see Sec.\ \ref{sec:neurologicalanalysis}).

For a human, $n$, at end $A$ we define $\tau_A^n$ (as illustrated in Fig.\ \ref{fig:tauAn}) as the time interval measured at end $A$ during which a human intervention has caused the setting at end $A$ to be changed and the latest subsequent time at end $A$ that a measurement could occur before a light speed signal could arrive at the corresponding measurement event at end $B$ carrying information from the location of human $n$ about this intervention.  During any such time interval, we will say that the human intervention is \emph{internal} (as it has influenced side $A$ but cannot have influenced side $B$).  We define $\tau_A$ as the average of $\tau^n_A$ taken over all the humans at end $A$.  We define the average $\tau_B$ for humans at end $B$ in a similar way.  We want $\tau_A$ and $\tau_B$ to be as big as possible. We will see that the effect scales as $\tau_A \tau_B$.

The humans who switch the setting at one end, $A$ say, should be on the opposite side of station $A$ to station $B$ (i.e. not between $A$ and $B$).  If we communicate the signal from these humans using radio frequencies then the $N_A$ humans who switch the setting at end $A$ could, in fact, be quite far from station $A$.  In particular, we could imagine the Bell experiment being performed in a remote location where conditions allowed for free space transmission or the easy laying down of optical fibers while the humans could be located in two towns or cities some kilometers away on either side of the Bell experiment.  The limiting factors here are: (i) the speed of light of radio frequency waves in air (typically this is about 1.0003 \cite{bean1959radio}), and (ii) the speed of the electronics used to transmit and receive radio frequency signals.   In the Canary Island experiment a radio frequency signal was successfully used to communicate the signal from the random number generator used to chose the setting at one end of the Bell experiment over a distance of 1.2km.

We define the delay at at ends $A$ and $B$ as
\begin{equation}
\tau_A^\text{delay} = \tau_A^\text{sep} - \tau_A   ~~~~~~~~~~~ \tau_B^\text{delay} = \tau_B^\text{sep} - \tau_B
\end{equation}
respectively.  Contributions to these delays come from: (i) delay getting the electrical signal from the interior of the brain to the EEG headset (this should be negligible): (ii) electrical delays in the equipment (this includes delays in any pre-processing of the EEG signal before it is inputted into the switch and delays in the device that switches the setting), (iii) delay in radio frequency signals used to connect headsets to switch due to transmission in air being slower than the speed of light; and (iv) geometric effects (if the humans are not on exactly behind the appropriate station).  These geometric effects will kick in if the humans are not on the line subtended from the line $AB$ behind the station they are giving input into.

\subsection{Neurological analysis}\label{sec:neurologicalanalysis}

Electroencephalograms were invented by Berger in the 1920's (see \cite{berger1929elektrenkephalogramm}).  Usually a power spectrum analysis is performed with the output in different frequency ranges being given names (for example, alpha is the frequency range 8-15Hz, beta is the frequency range 16-31Hz, \dots).  Different frequency ranges are associated with different mental functions and their analysis can help diagnose different pathologies. For our purposes we are only interested in using the signal as a source of candidate interventions.  We need to optimize our use and analysis of the signal with this goal in mind.

If we adopt the strategy of conditioning on certain EEG signals taken to be associated with human choice then we would need to develop an understanding of this.  There is an active field of research setting up a biological computer interface (BCI). Certain types of EEG signals are strongly correlated with different choices. For example, EEG analysis can predict which key on a computer keyboard is going to be pressed in typing tasks $\frac{1}{10}$th of a second before the key is actually pressed with greater than 96\% accuracy \cite{blankertz2002classifying}.  Since we are just interested in human \lq\lq interventions" that can be used to decide the setting in a Bell experiment, we are open to using signals associated with any kind of mental process.  If, for example, we suppose that such interventions are involved in, say, mathematical thinking then we could use EEG signals associated with this activity.

We can subject the EEG signals to a template analysis where these templates are supposed to pick out candidates for cases where there has been a human intervention.
For a given template choice, we are interested in the cases that some EEG brain activity passes this template test.  There is a question as to whether we (i) do this template analysis in real time (designing our apparatus to switch only when the EEG signal passes the given template) or (ii) simply use the EEG signal as the input into the device switching the setting and only subsequently do the template analysis which can be used to pick out the subensemble for which there are internal candidate interventions at both ends.  The first strategy would introduce additional delays into $\tau_{A,B}^\text{delay}$ so the second strategy seems better. Another advantage of the second strategy is that we can filter on different templates.  An advantage of the first strategy is that we can then more easily deduce the value of the retarded setting and directly evaluate the Bell inequalities with retarded settings discussed in Sec.\ \ref{sec:Bellretarded}.

Let $r_\text{human}$ be rate at which an individual human (on average) is able to communicate signals passing the given template test to the switching device of the Bell experiment. Since the relevant EEG activity seems to be at a frequency of order 10Hz, we will suppose $r_\text{human}=10$Hz.  Further, let the rate at which $N_A$ humans at end $A$ can communicate free choices to the switch be $r_A$ (and, similarly, at end $B$ we have $r_B$).  We suppose that $r_A=N_A r_\text{human}$ and $r_B=N_B r_\text{human}$ (strictly this will only be true up to a point as the electronics will fail to separately implement these switches for sufficiently high rates).

We may also want to design the EEG headsets to maximize the rate of such cases.  There are numerous EEG headsets available on the market now (many of them bluetooth enabled) for recreational use.  These are relatively cheap and easy for the wearer to put on.  However, it is not clear whether the electronics is fast enough for our purposes.  Also, we may gain a cost and logistical advantage using fewer humans with more effective headsets.  Since we want to optimize the use of these headsets with a different purpose in mind than the usual medical use of EEG headsets, it may make more sense to custom build them.

\subsection{Fast electronics and radio transmission}

We would need to develop fast electronics and radio transmission (if this is used) to get the signal from the brain activity to the switch at each end on a sufficiently fast time scale so that $\tau_{A,B}^\text{sep} > \tau_{A,B}^\text{delay}$.  Existing experiments \cite{weihs1998violation, scheidl2010violation, hensen2015loophole, giustina2015significant, shalm2015strong} have successfully implemented fast electronics from a random number generator.  The additional challenge here is to do this for EEG signals from a large number of humans.

\subsection{People management}

A new feature of such an experiment is that we would have to manage large numbers of people, fit them with headsets and engage them in some appropriate activity to maximize the value of $r_\text{human}$.   As we will see, the rate at which we can collect useful data scales as $N_AN_B$ so the payoff of having a large number of people is very significant.  It seems practical to have $N_A=N_B\approx 100$ for a time period on the order of an hour.  We could imagine being more ambitious and have a thousand people at each end for time on the order of a day but then the people management problems become significantly larger.  We could imagine arranging outreach events to attract people to attend. Or we could ask people to wear headsets during their normal work day.  Use of wireless EEG headsets (e.g. bluetooth) would be less of an imposition on the humans involved in this experiment.

\subsection{Data Analysis}

We would need to develop statistical methods to analyze the data from this type of experiment.  We could searching on different EEG signal templates for an effect.  Further, we could attempt to record the \emph{retarded settings} at each end (this will be discussed below) and use Bell inequalities with retarded settings.

Ideally, we would collect a data stream from each EEG headset as well as recording the setting at each end and also, of course, the outcome of the measurement. All of this information would have to be time stamped so we know what events are cross correlated.  It might be unmanageable to to collect this volume of data. In this case we would have to find ways to record, in real time, the salient information.  Recording all the data would allow subsequent data analysis. For example, we could investigate different candidate templates on the EEG signal. 

\subsection{Controls}

We can implement various controls on this experiment.  In the event that a violation of Quantum Theory were seen we would need, first, to see if there were some instrumental effect causing this anomalous result.  For example, the EEG signals passing our templates could induce a spurious behaviour of the apparatus that switches the measurement (after all, it is easy to introduce noise that degrades any violation of Bell's inequalities).  We could test for such effects by recording the EEG signals and running the experiment again using the recordings instead of humans.  Additionally, we could put a delay between the EEG headsets and the switching.  This could be used to set $\tau_A$ and $\tau_B$ to zero.  If the observed effect were really due to having humans switching while the systems were in flight then any effect ought to disappear when such recordings or when long enough delays are used.

\section{Fraction of cases with switching}

To be able to run a successful test we need to have a sufficiently large number of coincidence detection events for which a candidate human intervention has happened at each end and there has not been sufficient time for the new setting to have been communicated to the other end.  The fraction of coincidences, $\alpha$, for which this is the case is
\begin{equation}
\alpha = r_A r_B \tau_A \tau_B
\end{equation}
assuming that $r_A\tau_A<< 1$ and $r_B \tau_B << 1$ (so that the time intervals during which a candidate human intervention is internal do not overlap too much). If these ratios are close to one or bigger than 1 then $\alpha$ is close to 1.  The quantity $\alpha$ tells us how much harder it is to perform a Bell experiment in which locality conditions are imposed using human switching.  We can write this as
\begin{equation}
\alpha = N_A N_B r_\text{human}^2\tau_A \tau_B
\end{equation}
We see that having many humans at each end gives us a big payoff.

The coincidence rate of useful events (where a signal about neither candidate human intervention can have reached the other end) is
\begin{equation}
r_\text{coinc}^\text{human}= \alpha r_\text{coinc}
\end{equation}
where $r_\text{coinc}$ is the rate of coincidence detections for the Bell experiment when we do not worry about imposing locality conditions for free choices.

The time taken to collect significant statistics is
\begin{equation}\label{Thuman}
T_\text{exp}^\text{human} = \frac{T_\text{exp}}{\alpha} = \frac{T_\text{exp}}{N_A N_B r_\text{human}^2\tau_A \tau_B}
\end{equation}
(where $T_\text{exp}$ was defined earlier as the time taken to collect enough data to get a violation of Bell's inequality to some given significance in the raw Bell experiment).

\section{Feasibility}\label{sec:feasibility}

It is instructive to insert some numbers into equation \eqref{Thuman}.  We will put $N_A=N_B=100$ and $r_\text{human}=10 \text{Hz}$.  We assume that $\tau_{A,B} \approx \tau^\text{sep}_{A,B}$ (so delays are small).  In previous experiments involving fast switching (but not humans) it was claimed that this condition was met for the electronics \cite{weihs1998violation, scheidl2010violation, hensen2015loophole, giustina2015significant, shalm2015strong}.  We will take $\tau^\text{sep}_{A,B}$ and $T_\text{exp}$ from three previous experiments to assess feasibility.  The relevant data from these experiments is shown in Fig.\ \ref{fig:experiments}.  The bottom line is to know the time, $T_\text{exp}^\text{human}$, it would take to perform a statistically significant experiment with humans doing the switching.

An experiment reported in 1997 had the source in Geneva \cite{tittel1998violation} and the two ends in the neighbouring towns of Bellevue and Bernex (a distance of 10.9km apart).  This experiment did not implement switching but it provides some numbers we can use to look at feasibility.  It took 20s to collect each data point.  After analyzing data from hundreds of such data points and performing a best fit, a violation of Bell's inequalities of about 10 standard deviations was seen. With $N_A=N_B=100$ and $r_\text{human}=10\text{Hz}$ we get $\alpha\approx10^{-3}$ for $d_\text{sep}=10.9\text{km}$. Thus we would require about 5 hours per data point in an experiment with humans. Hundreds of such data points were collected in this experiment.  In principle, we only require four data points (as we have two settings at each end) so this would require about 20 hours.  However, that the 10 standard deviations mentioned above was a best fit with hundreds of data points giving $T_\text{exp}\approx 1~\text{hour}$.  Consequently, we would actually require $T_\text{exp}^\text{human}\approx 1000~\text{hours}\approx 40~\text{days}$.

The 1998 Innsbruck experiment \cite{weihs1998violation} had $d_\text{sep}=400\text{m}$.  With the above choices we obtain $\alpha\approx 10^{-6}$.  In this experiment a violation of Bell's inequalities was observed to 30 standard deviations using data (comprising about 15000 coincidences) collected over only 10s.  A similar experiments with humans would require $10 \times 10^{6}$ seconds per data point (or about 4 months). It is clearly not feasible to run an experiment like this with 200 people in place for 4 months.

In the 2010 experiment performed between in the Canary Islands between La Palma and Tenerife \cite{scheidl2010violation}, the distance between the two stations was $144\text{km}$.  The source was located at La Palma where one photon from each entangled pair was sent through a $6\text{km}$ coil of optical fibre before measurements and the other photon was transmitted through $144\text{km}$ of free space to a telescope located at Tenerife where measurements were made. The distance between the two ends of the experiment in the frame of reference in which the measurements were simultaneous was calculated to be $50\text{km}$.  By considering the geometry of the setup we have $c\tau_A=6\text{km}$ and $c\tau_B=144\text{km}$.  With 100 humans at each end and a rate, $r_\text{human}=10\text{Hz}$, as above we get $\alpha\approx 0.01$.  This is clearly much better as we would only have to collect data for about 100 times as long to get similarly significant figures.  In this experiment it took 600s to collect enough data for a violation of Bell's inequalities of 16 standard deviations so we should be able to get a similar significance in 16 hours.

\begin{figure}
\begin{tabular}{|l|| c | c || c | c|}
\hline
Experiment  &  $d_\text{sep}$ & $T_\text{exp}(\text{significance})$ & $\alpha$ & $T_\text{exp}^\text{human}(\text{significance})$  \\
\hline\hline
Geneva 1997 & 10.9km & 1 hour(10) & $10^{-3}$ & 40 days(10) \\
\hline
Innsbruck 1998&    400m  &   10s(30) & $10^{-6}$ & 4 months(30) \\
\hline
Canary 2010 & 50km & 10 min(16) & $10^{-2}$ & 16 hours(16) \\
\hline
\end{tabular}
\caption{The first three columns of this table provide information pertaining to three experiments: The 1997 Geneva experiment \cite{tittel1998violation}, the 1998 Innsbruck experiment \cite{weihs1998violation}, and the 2010 Canary Islands experiment \cite{scheidl2010violation}. The distance between the two ends is $d_\text{sep}$ and time taken to collect the data is $T_\text{exp}(\text{significance})$ where the significance is given in parenthesis as the number of standard deviations violation of Bell's inequalities.  The last two columns provide  an estimate for the faction $\alpha$ is provided (with 100 humans at each end assuming $r_\text{human}=10Hz$) and the time taken, $T_\text{exp}^\text{human}(\text{significance})$, to perform an experiment with humans providing the switching (with the number of standard deviations given in parenthesis again).}
\label{fig:experiments}
\end{figure}

Sixteen hours is, perhaps, too long to keep 200 people wearing EEG headsets (in addition to keeping the experiment stable). However, it is not too far from being feasible.  If we accept a lower significance then we could get a result more quickly. The error scales as $\frac{1}{\sqrt{N}}$ where where $N$ is the number of coincidence counts. Hence, the number standard deviations seen scales as $\sqrt{N}$ which scales as $\sqrt{\frac{T}{\alpha}}$ where $T$ is the time we collect data.   Then, all other things being equal, we can get a violation of the Bell inequalities (with the human switching locality condition imposed) of about 6 standard deviations in about two hours (or, a similarly significant violation of quantum theory if the effect we are looking for actually exists with the given value of $r_\text{human}$).  Of course, many other improvements could be made. For example, in recent years photon detectors with much higher efficiencies have been built.  The coincidence rate depends on the square of the detector efficiency so this could make a big difference.  We could also imagine a more symmetric geometry in which a central source is viewed by two telescopes each at a distance on the scale of 100km.  Additionally, we could consider using more people at each end.

These feasibility calculations all used $r_\text{human}=10\text{Hz}$.   The main justification for this is that this is, roughly, the frequency of EEG signals.  Even in the case that humans can make interventions that would violate Quantum Theory in a Bell experiment, it is possible that $r_\text{human}$ is much smaller.  Since the effect scales as $r_\text{human}^2$ we are particularly sensitive to this.  It is also possible that $r_\text{human}$ is bigger than 10Hz (there is fine grained structure on the EEG scans at higher frequencies than 10Hz).

One advantage of this experiment is that we can use the massive volume of data that is collected even when the locality conditions are not satisfied (i.e.\ when there has not been a human induced switch at each end) to stabilize the experiment.  This data will violate Bell's inequalities.

\section{Bell inequalities with retarded settings}\label{sec:Bellretarded}

In \cite{hardy1989proposal, hardy1989local, hardy2017bell} I derived Bell inequalities with retarded settings for the purpose of this type of experiment.  The idea is to take into account the retarded values of the settings.  The retarded value, $b_r$, of the setting at end $B$ as seen at end $A$ is the value the setting at end $B$ is predicted to take on the basis of  information that can be locally communicated to $A$ (see Sec.\ \ref{sec:inflight} for more details). If there is a human intervention changing the setting at end $B$ happening after the time a signal could have been sent from end $B$ to end $A$, then the actual setting, $b$, and retarded setting, $b_r$, will be different.  We can define $a$ and $a_r$ as the actual and retarded settings at end $A$ similarly.

We assume that the outcome at end $A$ can depend on $a$, $a_r$, $b_r$, and some local hidden variables (but not on $b$). We write this as
\begin{equation}
A(a, a_r, b_r, \lambda)
\end{equation}
where this is the result of the measurement.  Similarly, at end $B$ we have
\begin{equation}
B(b, b_r, a_r, \lambda)
\end{equation}
We will assume that the results at each end are equal to $\pm1$. Here $\lambda\in \Gamma$ is a hidden variable with some distribution $\rho(\lambda)\geq 0$ such that
\begin{equation}
\int_\Gamma \rho(\lambda) d\lambda = 1
\end{equation}
We define the correlation function
\begin{equation}
E(a,b|a_r, b_r) = \int_\Gamma  A(a, a_r, b_r, \lambda) B(b, b_r, a_r, \lambda) d\lambda
\end{equation}
We use the mathematical result that
\begin{equation}\label{CHSHmathinequality}
 X'Y' +X'Y + XY' -XY =\pm 2
\end{equation}
where $X, X', Y, Y' =\pm 1$. We put
\begin{align}
X &= A(a, a_r, b_r, \lambda)\\
X'&=A(a', a_r, {b}_r, \lambda)\\
Y &=B(b, a_r, b_r, \lambda) \\
Y'&=B(b', {a}_r, b_r \lambda)
\end{align}
If we substitute these into \eqref{CHSHmathinequality} and integrate over $\lambda$ we obtain
\begin{equation}\label{CHSHretarded}
-2 \leq E(a',b'|{a}_r, {b}_r)+E(a',b|{a}_r, {b}_r)+E(a,b'|{a}_r, {b}_r)-E(a,b|a_r, b_r) \leq +2
\end{equation}
These are the Bell inequalities with retarded settings in Clauser Horne Shimony Holt form (i.e. based on the CHSH  \cite{clauser1969proposed} of the Bell inequalities).  We can also also obtain such inequalities in Clauser Horne \cite{clauser1974experimental} form (see \cite{hardy1989proposal, hardy1989local, hardy2017bell}).  Note that (unlike in previous papers \cite{hardy1989proposal, hardy1989local, hardy2017bell}) we have allowed the result at end $A$ to depend on the retarded setting, $a_r$ at this end (and similarly at end $B$).

It is shown in \cite{hardy2017bell} that it is easy to write down a local model (in which dependence on retarded values is allowed) that reproduces the predictions of Quantum Theory when the actual and retarded settings are equal.  The model works as follows.
Let
\begin{equation}
0\leq \lambda < 2\pi ~~~~~~~ \rho= \frac{1}{2\pi}
\end{equation}
Where $\rho(\lambda)$ is the distribution function (so we have a flat distribution). Define
\begin{equation}
A(a, b_r, \lambda) = \left\{ \begin{array}{ll}
                             +1 & \text{for}~ \theta_L \leq \lambda < \theta_L +\pi \\
                             -1 & \text{for}~ \theta_L +\pi \leq \lambda < \theta_L+2\pi
                             \end{array}  \right\}
\end{equation}
and
\begin{equation}
B(b, a_r, \lambda) = \left\{ \begin{array}{ll}
                             +1 & \text{for}~ \theta_R \leq \lambda < \theta_R +\pi \\
                             -1 & \text{for}~ \theta_R +\pi \leq \lambda < \theta_R+2\pi
                             \end{array}  \right\}
\end{equation}
It is easy to prove that
\begin{equation}
E(a, b|a_r,b_r) = 1-\frac{2|\theta_R-\theta_L|}{\pi}
\end{equation}
If we set
\begin{equation}
\theta_L= -\frac{\pi}{4}(1+\cos(a-b_r))   ~~~~ \theta_R = \frac{\pi}{4}(1+\cos(a_r-b))
\end{equation}
we obtain
\begin{equation}
E(a,b|a_r,b_r) = -\frac{1}{2}(\cos(a-b_r) + \cos(a_r-b))
\end{equation}
When the retarded settings are equal to the actual settings we get
\begin{equation}
E(a,b|a,b)=-\cos(a-b)
\end{equation}
in agreement with Quantum Theory.  Note, incidentally, that with this model we get a violation of Quantum Theory even when the actual and retarded settings are different for only one end. This is not demanded by the Bell inequalities with retarded settings.  It should be possible to build a model that gives the quantum predictions as long as the actual and retarded settings are equal for at least one of the two ends.

If we have some candidate as to what counts as an intervention then we can, in principle, determine the value of the retarded settings.  Then we can test the retarded inequalities directly.  We can only be sure that a local hidden variable model will violate Quantum Theory when the actual and retarded settings are different at each end.  It is, however, worth investigating what happens when this locality condition is only obtained at one end. Indeed, it is much easier to implement these conditions experimentally.  Furthermore, as we have just seen, there do exist models having the property that Quantum Theory is violated even when the actual and retarded settings are different for only one end.

One way to be sure that we can determine the retarded settings is to only allow the setting to be changed by those EEG signal that have been identified as candidate interventions (these are our EEG signal templates).  Then the retarded setting is equal to the setting at the earlier time (intersected by the past light cone from the measurement event on the other side).  However, this requires us to put electronics in place up front that can implement this conditional switching.  Such electronics would introduce additional delays reducing the overall effect we see.  Further, this would only allow us to investigate one candidate EEG signal template. It is better to allow the all of the EEG signal to switch the settings.  Then we can analyze the data for different candidate EEG signal templates.  We can only expect to marshal hundreds of people to participate in this experiment for limited time and so it is better to be able to investigate multiple EEG signal templates.

If we are not able to determine the retarded settings then we should think about whether the distribution over them is even or not.  In \cite{hardy2017bell} it is shown how to go from Bell inequalities with retarded settings to standard Bell inequalities under certain assumptions about the distribution over the retarded settings.  Failure to investigate this might allow local hidden variable models that take into account the retarded settings to produce a violation of the standard Bell inequalities.  An extreme example of this type is the original Aspect et al experiment \cite{aspect1982experimental} which used periodic rather than random switching \cite{hardy2017bell,zeilinger1986testing}.

\section{Connection to mind matter debate}\label{sec:mindmatter}

\subsection{Consciousness}

Humans are conscious.  There is no generally agreed scientific account of this.  One viewpoint is that consciousness is something that will emerge in systems that are sufficiently complex in the right way.  For example, a computer program that is functionally indistinguishable from a human (as in Alan Turing's \lq\lq imitation game" \cite{turing1950computing}) might be argued to be conscious. Another possibility is that certain sorts of physical process give rise to consciousness.  Roger Penrose has suggested that quantum superposition may play a role in mental processes \cite{penrose1991emperor}.  The idea is that objective reduction events (collapses of the wave function) in the brain demanded from arguments from Quantum Gravity are non-computable and provide the missing link to explain consciousness. For such quantum effects to play a role in cognition, we would need quantum superposition to be possible in the brain (otherwise we would expect a fully classical account to be possible). Penrose and Hameroff proposed that quantum superpositions can be maintained in microtubules in the brain \cite{hameroff1996orchestrated}.  Others have weighed in on this debate. Max Tegmark argued that decoherence effects are too strong in the brain to allow quantum superposition in microtubules \cite{tegmark2000importance}.  Matthew Fisher proposed, instead, that nuclear spins could maintain quantum superpositions allowing quantum effects to play a role in cognition \cite{fisher2015quantum}.

It is difficult to conceive of how something as distinctive and singular as consciousness could emerge once something becomes complex.  Similarly, it is hard to see how consciousness could be a result of certain types of physical process. Why, for example, would non-computability bring about the sensation of being conscious?
Why are some systems (humans for instance) capable of having experiences? This was called the \lq\lq hard problem of consciousness" by David Chalmers \cite{chalmers1995facing}.

\subsection{Cartesian dualism}

One viewpoint is that that consciousness is due (at least in part) to some sort of non-physical mind outside regular physics that is endowed with the property of consciousness.  This is the Cartesian duality point of view put forward by Descates in 1641 (translated in \cite{descartes1985meditationes}). As more and more aspects of human mental functioning have been accounted for by studying the brain, the case for this kind of dualism has receded.  Furthermore, dualism has echoes of a pre-scientific attitude towards nature that is now largely discredited in scientific circles.  However, if we were able to make some empirical progress of the sort suggested in this paper then these criticisms would be mute - experiment is, after all, the final arbitrator in science.

Even without taking into account issues arising from Bell's theorem, dualism ought to have some empirical consequences.  An argument for this in \cite{squires1990conscious} is to note that the word \lq\lq consciousness" appears in ink in the dictionary.  The fact that the atoms in the ink got into that particular configuration must have been influenced by whatever is responsible for the attribute of consciousness. A community of non-conscious robots would, presumably, not invent and consistently use the term \lq\lq consciousness". In other words, there seems to be little support for a notion of duality in which the minds merely passively observe without influencing matter.

From the dualistic viewpoint we can tell a story in which the mind acts on the brain and thereby imparts information into the physical world (that, somewhere down the line, can lead to certain configurations of ink in a dictionary).  We could imagine looking at the detailed behavior of atoms inside the brain searching for a violation of the standard rules of physics.  However, it is not really clear what we would be looking for and, besides, it would be very hard to verify a violation of Quantum Theory (or whatever the prevailing scientific theory is) in such a messy environment.  The proposal described in this paper offers a way forward without having to make any kind of precision measurements inside the brain.

From our present point of view such \lq\lq mind-acts" provide candidate interventions that could be used to determine the settings in a Bell experiment. The important point is not whether humans have free will as such, but rather whether the effects of mind on matter can be \lq\lq anticipated" by the laws of physics.

We will describe two types of model that would lead to a violation of Quantum Theory in accord with Bell's inequalities when humans are used to determine the settings.  These are \emph{local (super)-deterministic dualistic theories} (LDD) \cite{hardy1989proposal, hardy1989local} and \emph{local (super)-deterministic interventionist brain theories} (LDFB).  It is not entirely clear that the second type of theory can be consistently formulated. We will discuss this.

Although there has been no discussion of mind-matter duality in the context of the Bell experiment (aside from \cite{hardy1989proposal, hardy1989local, hardy2017bell}) there has been much discussion of it in other areas of Quantum Foundations - in particular in attempted resolutions of the measurement problem.  In 1932 von Neumann \cite{von1955mathematical} argued that it makes no difference whether the projection postulate is applied on the system immediately prior to measurement, at the measurement apparatus, or at the level of the brain (or anywhere in between).  In 1939 London and Bauer proposed that it happens at the point when observer becomes conscious of the measurement outcome (their article is reprinted in \cite{wheeler2014quantum}). Wigner pushed this point of view also coming up with his well known Wigner's friend example.  In more recent years, Stapp has developed a point of view along these lines \cite{stapp2004mind}.   A variant of the many worlds interpretation is the many minds interpretation \cite{zeh1970interpretation, albert1988interpreting,lockwood1996many, donald1999progress}.  Gisin asserted that the Copenhagen interpretation is dualistic in nature \cite{gisin2017collapse}.

The application of dualism proposed in this paper is quite different.  In particular, the proposal in this paper does not relate to the measurement problem of Quantum Theory as such.  Rather, it is an attempt to provide interventions in determining the settings of a Bell experiment.

While the existence of consciousness is motivation to posit dualism, it is not so clear that we can claim that dualism \emph{explains} consciousness. In the end, in science, we are only able to offer a place for the most fundamental properties we see (less fundamental properties are explained in terms of these more fundamental properties).  Dualism, at least, offers place for consciousness understood as a fundamental property.

\subsection{Super-deterministic theories}

Before we bring discussion of humans and minds into the picture we want to recap the idea of super-determinism.  It is well known that it is possible to maintain locality in the face of the observed violations of Bell's inequalities if we have determinism \cite{hardy1989proposal, hardy1989local, sarlemijn2012nouvelle, scheidl2010violation, t2014cellular}.  The idea being that the settings at each end are taken to be determined by conditions in the past light cone of both ends. Hence the information about the setting can be communicated to the other end without violating locality. This is usually called super-determinism \cite{sarlemijn2012nouvelle}. One way to do this is to (extravagantly) imagine that information about the initial state of the entire universe is encoded into every spacetime point.

We could attempt to guard against this super-determinism by performing experiments wherein signals from causally disconnected parts of the universe are used to determine the settings at each end (see discussion in Sec.\ \ref{sec:inflight}).

\subsection{Local (super)-deterministic dualistic theories}

Now we introduce humans into the picture.  We suppose some sort of mind-matter duality in which the physical universe is local and super-deterministic. In the absence of minds, then, it is possible to violate Bell inequalities.  Now we suppose that minds act on the physical universe locally introducing \lq\lq new information" into the physical universe through the brain. This information spreads out locally such that, one second after such a mind-act, this new information can be available $3\times 10^8$m away.  It is clear that we can derive Bell inequalities for this kind of situation in which humans choose the settings at each end (so long as we ensure that the new information cannot have reached the measurement event at the other end).  We call theories of this kind local (super)-deterministic dualistic theories (LDD).  Such theories were originally put forward in \cite{hardy1989proposal, hardy1989local}.

\subsection{Local (super)-deterministic interventionist brain theories}

To examine whether we would be forced to dualism were Quantum Theory violated (in accord with Bell's inequalities) when humans are used to choose the settings, we can attempt to outline a non-dualist class of theories that would also lead to this result.   It could be the case that physics is super-deterministic except for systems that are complex in the kind of way that happens inside brains.  Were the world described by such a theory then settings chosen by humans will count as interventions.  We call such theories local (super)-deterministic interventionist brain theories (LDIB).

It is difficult to see how to build these type of theories especially if we want to have local microphysical laws (a reasonable demand if we are interested in constructing a local theory).  For then the local microphysical laws would have to apply to small collections of atoms in the brain as they apply to small collections outside.  For outside atoms these laws would have to be super-deterministic so as to account for already performed Bell experiments in a local way.  But theories with local microphysical laws satisfy reductionism - we can account for the behavior of a composite system in terms of the behavior of its parts. But then it would seem that brains, taken as a whole, should be super-deterministic.  This is not a rigorous proof against the possibility of LDIB theories but it does point to some serious difficulties with constructing them.

If we imagine it were established fact that we only get a violation of Quantum Theory in accord with Bell's inequalities when humans (and, presumably, other animals) are used to choose the settings then we would have to explain this in terms of local physics.  In particular, we would have to explain why (in this imagined scenario) other similarly complex physical systems do not produce a similar violation of Quantum Theory.

We could turn this round and attempt to outline a theory in which only systems with certain types of complexity lead to the ability to make interventions and that these become natural candidates for conscious systems.  An analogue for this idea is to imagine the surface of a lake where ripples and waves follow the usual laws of wave physics. However, a fish poking its nose at the surface from below, or a bird skimming the surface from above could not be anticipated by the laws of wave physics and would appear as interventions. In this analogue, however, we will not see a violation of Bell inequalities as such.  Consequently, much of the motivation for the foregoing discussion is absent.  Further, robotic fish and flying drones would produce the same effects as fish and birds in this system whereas the hypothetical result we are considering is wherein humans (and other animals) produce different results than robots, random number generators, and other machines.

It is doubtful that we can consistently build LDIB theories for the above reasons.  The main problem is that there may be breakdown of reductionism.  If we consider extracting our random choices from just a few atoms (inside or not inside a brain) then this, surely, would not lead to a violation of Quantum Theory in a Bell experiment.

\subsection{Turing-style test}

\begin{figure}
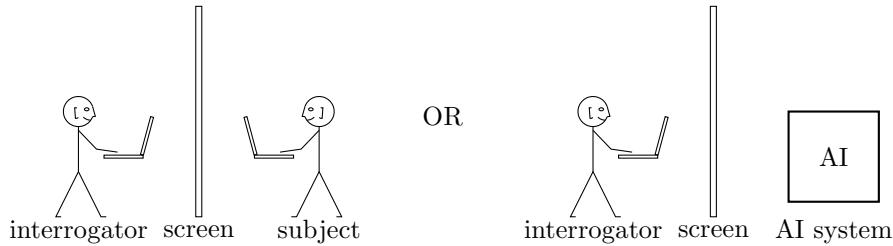

\begin{Compose}[0.8]{0}{0}
\humanforward[1]{0,0}\laptop{0,0} \Thistexthere{interrogator}{0,-8}
\crectangle[thin]{Screen}{0.2}{7}{8,0} \Thistexthere{screen}{8,-8}
\humanforward[1, xscale=-1]{16,0} \laptop[1, xscale=-1]{16,0}  \Thistexthere{subject}{16,-8}
\end{Compose}
~~~~~OR~~~~~
\begin{Compose}[0.8]{0}{0}
\humanforward[1]{0,0}\laptop{0,0} \Thistexthere{interrogator}{0,-8}
\crectangle[thin]{Screen}{0.2}{7}{8,0} \Thistexthere{screen}{8,-8}
\crectangle[thick]{AI}{3}{3}{16,-3} \csymbol{\ensuremath{\text{AI}}} \Thistexthere{AI system}{16,-8}
\end{Compose}
\caption{The Turing test.  The interrogator asks questions over a computer interface to determine whether he is talking to another human (as shown in picture on the left) or an artificial intelligence system (as shown in the picture on the right).}\label{fig:Turingtest}
\end{figure}

Turing was interested in the question of whether programable computers can simulate humans.  He proposed a test \cite{turing1950computing} in which a human interrogator can ask questions to a mystery system.  This mystery system is either a human or some artificially intelligent device (a computer with a suitable program).  The interrogator's questions are communicated by a keyboard and he reads the answers off a screen.   The objective for the interrogator is to guess whether he is talking to a human or an artificial intelligence device (see Fig.\ \ref{fig:Turingtest}).  If such artificial intelligence devices can consistently fool human interrogators then we have evidence that human brains are effectively equivalent to a computer program.  This would provide evidence against mind matter dualism - if a computer is functionally equivalent to a human then what need is there for some sort of non-physical mind?  The Loebner Prize is an annual competition running since 1991 to write a computer program that can pass the Turing test.  This has shown a steady improvement in the ability of computers to fool human interrogators \cite{loebnerprize}. However, humans still outperform computers in these tests.

\begin{figure}
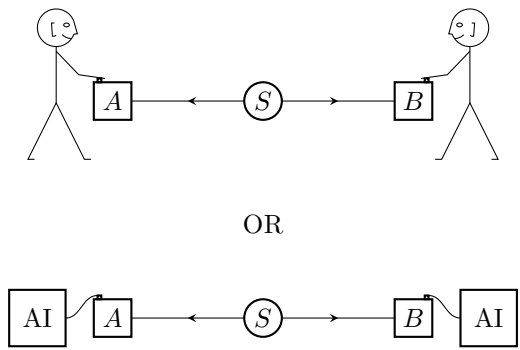

\begin{center}
\begin{Compose}{0}{0}
\Ccircle{S}{1}{0,0}
\Csquare{A}{1}{-8,0} \csquare{buttonA}{0.1}{-8.7,1.1}
\Csquare{B}{1}{8,0}  \csquare{buttonB}{0.1}{8.7,1.1}
\joincr{S}{180}{A}{0}
\joincl{S}{0}{B}{0}
\humanforward{-11,3.9}
\humanforward[1, xscale=-1]{11,3.9}
\end{Compose}
\end{center}
\vskip 2.5mm
\begin{center}{OR} \end{center}
\vskip 2.5mm
\begin{center}
\begin{Compose}{0}{0}
\Ccircle{S}{1}{0,0}
\Csquare{A}{1}{-8,0} \csquare{buttonA}{0.1}{-8.7,1.1}
\Csquare{B}{1}{8,0}  \csquare{buttonB}{0.1}{8.7,1.1}
\joincr{S}{180}{A}{0}
\joincl{S}{0}{B}{0}
\csquare{AI}{1.5}{-12,0} \csymbol{\ensuremath{\text{AI}}} \joinrlnoarrow{AI}{0}{buttonA}{0.1}
\csquare{AI}{1.5}{12,0}  \csymbol{\ensuremath{\text{AI}}} \joinlrnoarrow{AI}{0}{buttonB}{0.1}
\end{Compose}
\end{center}
\caption{A Turing style test but using a Bell experiment.  Pairs of humans compete against pairs of AI systems.}
\label{fig:TuringBelltest}
\end{figure}

One problem with the Turing test is that it depends on the subjective judgement of interrogator.  Here we propose an alternative test, utilizing a Bell experiment, to distinguish humans from artificial intelligence devices.    Now we need two humans or two AI devices (or whatever type of system we want to use to compete with humans -the \lq\lq I" in AI may more appropriately stand for \lq\lq intervention").  The role of the interrogator is now played by the Bell experiment.  The two humans or two artificial intelligence devices are placed so as to provide input to switches at each end of the Bell experiment (see Fig.\ \ref{fig:TuringBelltest}).  The objective is to cause a Bell experiment (which under ordinary operation, violates Bell inequalities by a significant amount) to satisfy the Bell inequalities only by providing the input to this switching.  In the case that the world is described by a local (super)-deterministic dualistic theory, humans will be able to do this but artificial intelligence devices will not. Unlike the Turing test, this test is objective. Were humans able to pass the test while artificial intelligence machines were not then this would provide evidence for mind-matter dualism.

\section{Discussion}

Quantum Theory has been tremendously successful empirically.  It seems very unlikely that a Bell experiment using humans as described in this paper, this would lead to a violation of Quantum Theory.  On my more optimistic days I would put the probability at about $1-2\%$ though I suspect many of my colleagues would give a much lower figure.  To have a violation of Quantum Theory in agreement with Bell inequalities under these circumstances is best motivated if we have (i) mind-matter duality and (ii) super-determinism.  Personally I think that we need some radical change to the scientific world view to deal with the hard problem of consciousness and so I would give a relatively high weighting to dualism (perhaps $30\%$). Super-determinism, on the other hand, would be a step back scientifically.  I prefer to think that the violation of Bell inequalities as seen in experiments so far is telling us something deep about causal structure in nature (that we will, perhaps, understand when we have a theory of Quantum Gravity).  If the explanation turns out to be super-determinism then the message of Bell's theorem will not be so deep after all.  Even if the probability of seeing a violation of Quantum Theory under these circumstances is much lower, there is still a very high
\begin{equation}
(\text{probability})\times (\text{payoff})
\end{equation}
Indeed, the payoff, in scientific terms, would be tremendous both in terms of importance for our understanding of Quantum Theory and, even more significantly, our understanding of mind.

Even if this experiment does not lead to a violation of quantum theory it shows how we can talk scientifically about mind-matter dualism.  There exists a class of scientifically testable theories invoking duality that are open to falsification.

\section*{Acknowledgements}

I am especially grateful to Mike Lazaridis for suggesting I write up a proposal for this experiment.  This paper is a direct consequence of that discussion.  I am also grateful to Hilary Carteret, Adrian Kent, Matthew Leifer, Markus Mueller, Rob Spekkens, Steven Weinstein, and Elie Wolfe for discussions.  I would like to thank the staff at Perimeter Institute Black Hole Bistro where a substantial fraction of this work was done.

This project/publication  was made possible through the support of a grant  from the John Templeton Foundation. The opinions expressed in this publication are those of the author(s) and do not necessarily reflect the views of the John Templeton Foundation.

Research at Perimeter Institute is supported by the Government of Canada through the Department of Innovation, Science and Economic Development Canada and by the Province of Ontario
through the Ministry of Research, Innovation and Science.

\bibliography{BellBib2}
\bibliographystyle{plain}

\end{document}